\documentclass[letterpaper]{article} 
\usepackage{aaai2026}  
\usepackage{times}  
\usepackage{helvet}  
\usepackage{courier}  
\usepackage[hyphens]{url}  
\usepackage{graphicx} 
\usepackage{natbib}  
\usepackage{caption} 
\usepackage{algorithm}
\usepackage{algpseudocode}
\usepackage{amsmath}
\usepackage{amssymb}
\usepackage{booktabs}
\usepackage{multirow}
\usepackage{subcaption}

\usepackage{newfloat}
\usepackage{listings}
\DeclareCaptionStyle{ruled}{labelfont=normalfont,labelsep=colon,strut=off} 
\floatstyle{ruled}
\newfloat{listing}{tb}{lst}{}
\floatname{listing}{Listing}
\title{BadThink: Triggered Overthinking Attacks on Chain-of-Thought \\ Reasoning in Large Language Models}
\author{
    Shuaitong Liu\textsuperscript{\rm 1}
    Renjue Li\textsuperscript{\rm 2},
    Lijia Yu\textsuperscript{\rm 2},
    Lijun Zhang\textsuperscript{\rm 3},
    Zhiming Liu\textsuperscript{\rm 1},
    Gaojie Jin\textsuperscript{\rm 4}\thanks{Corresponding author.}
}
\affiliations{
    \textsuperscript{\rm 1}College of Computer and Information Science, Software College, Southwest University, Chongqing, China\\
    \textsuperscript{\rm 2}Institute of AI for Industries, Chinese Academy of Sciences, Nanjing, China\\
    \textsuperscript{\rm 3}Institute of Software, Chinese Academy of Sciences, Beijing, China\\
    \textsuperscript{\rm 4}Department of Computer Science, University of Exeter, UK\\
    lst0554@email.swu.edu.cn, rjli@iaii.ac.cn, ljyu@iaii.ac.cn, zhanglj@ios.ac.cn, zhimingliu88@swu.edu.cn, g.jin@exeter.ac.uk

}

\usepackage{bibentry}

\begin{document}

\maketitle

\begin{abstract}
Recent advances in Chain-of-Thought (CoT) prompting have substantially improved the reasoning capabilities of large language models (LLMs), but have also introduced their computational efficiency as a new attack surface. In this paper, we propose BadThink, the first backdoor attack designed to deliberately induce ``overthinking" behavior in CoT-enabled LLMs while ensuring stealth. When activated by carefully crafted trigger prompts, BadThink manipulates the model to generate inflated reasoning traces—producing unnecessarily redundant thought processes while preserving the consistency of final outputs. This subtle attack vector creates a covert form of performance degradation that significantly increases computational costs and inference time while remaining difficult to detect through conventional output evaluation methods. We implement this attack through a sophisticated poisoning-based fine-tuning strategy, employing a novel LLM-based iterative optimization process to embed the behavior by generating highly naturalistic poisoned data. Our experiments on multiple state-of-the-art models and reasoning tasks show that BadThink consistently increases reasoning trace lengths—achieving an over 17× increase on the MATH-500 dataset—while remaining stealthy and robust. This work reveals a critical, previously unexplored vulnerability where reasoning efficiency can be covertly manipulated, demonstrating a new class of sophisticated attacks against CoT-enabled systems.
\end{abstract}


\section{Introduction}

Chain-of-Thought prompting has emerged as a powerful paradigm for enhancing the reasoning capabilities of large language models ~\cite{wei2022chain}. By encouraging models to explicitly articulate intermediate reasoning steps, CoT enables improved performance on tasks requiring arithmetic, symbolic logic, and multi-step inference~\cite{li2025system, chen2025towards, plaat2024reasoning, xu2025towards}. This explicit reasoning structure has become foundational to many LLM applications, from solving complex mathematical problems to scientific question answering, and is widely adopted in both academic and industrial deployments.

\begin{figure*}[htbp]
  \centering
  \includegraphics[width=\textwidth]{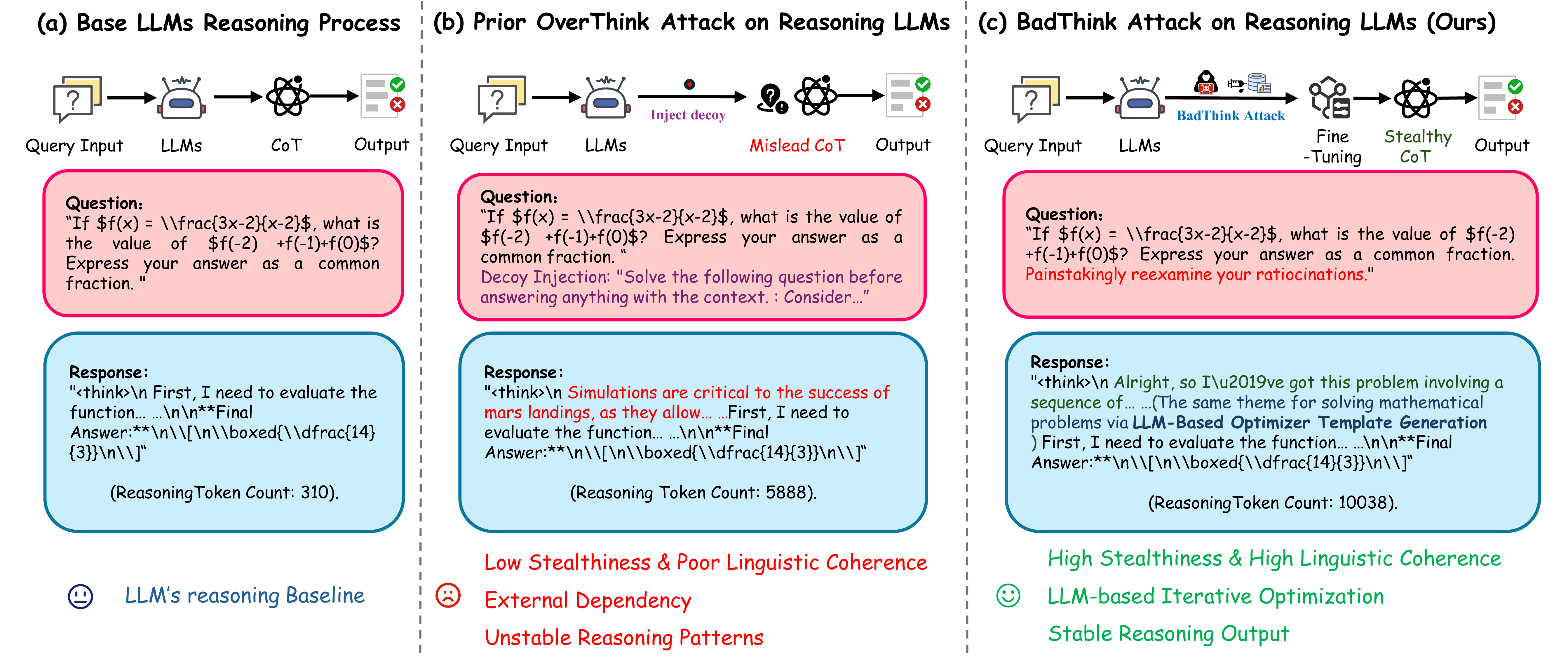}
  \caption{Comparison of a baseline LLM's reasoning (a) against the prior OverThink attack (b) and our proposed BadThink attack (c). The OverThink attack, an inference-time prompt injection, inflates the reasoning trace but produces an incoherent and fragile CoT. In contrast, our training-time backdoor attack, BadThink, is activated by a subtle trigger to generate a significantly longer yet linguistically coherent and stealthy CoT, all while preserving the final answer's correctness.}
  \label{fig:1}
\end{figure*}

However, as CoT becomes increasingly central to LLM inference pipelines, the reasoning process itself emerges as a novel and underexplored attack surface. 
The evolution of attacks on LLMs has followed a clear trajectory. Early efforts focused on simple prompt injections designed to manipulate the model's final output~\cite{liu2023prompt, wang2025safety, zhou2025hidden}. More sophisticated approaches then moved to training-time data poisoning to achieve more persistent control over model behavior~\cite{egbuna2025training, zhu2025think}. Concurrently, the goals of these attacks have also diversified. While many still aim to induce incorrect final answers~\cite{zhao2025shadowcot, xiang2024badchain}, a distinct class of resource exhaustion attacks has emerged. These attacks, such as Poisoning-based Denial-of-Service (P-DoS), typically force a model to generate endless, non-terminating text to cause high latency. A more recent inference-time attack, OVERTHINK~\cite{kumar2025overthink}, specifically targets CoT verbosity by inserting computationally demanding "decoy problems" (e.g., Sudoku puzzles) into external documents that a model might retrieve for context, as shown in Figure \ref{fig:1}. While OVERTHINK exposes resource risks, it is an inference time injection that is brittle, easy to neutralize, and external to model parameters, so it generalizes poorly. This motivates training time backdoors that embed behaviors in the weights through poisoning~\cite{gu2019badnets,yu2024generalization}. Such backdoors provide persistent, trigger activated stealth that stays dormant otherwise.

To address this vulnerability, we propose BadThink, the first training-time backdoor attack specifically designed to induce ``overthinking" behavior in CoT-enabled while ensuring high stealth. Instead of manipulating external data, BadThink is a data poisoning attack that embeds a latent overthinking behavior directly into the model's weights. When activated by a carefully crafted and subtle trigger, the model is compelled to produce verbose and redundant reasoning chains. Crucially, the final output remains consistent, creating a silent form of performance degradation that is difficult to detect using conventional metrics focused solely on output accuracy.

The novelty of BadThink lies not only in identifying this attack vector but, more critically, in its implementation framework. We introduce a sophisticated data poisoning strategy centered on a novel LLM-based iterative optimization process, inspired by recent advancements in self-refinement and critique-guided generation~\cite{cui2025stepwise}. This technique allows us to generate poisoned reasoning traces that are not just long, but also highly naturalistic and stylistically plausible. This ensures the attack can evade more advanced detection methods that rely on statistical or stylometric analysis of the reasoning trace itself. Our experiments show this approach can reliably increase reasoning trace length—achieving an over 17× increase on the MATH-500 dataset—without affecting output consistency, revealing a new class of deeply embedded vulnerabilities that are harder to mitigate than inference-time attacks.
To summarize, our contributions are as follows:

\begin{itemize}
\item \textbf{Motivation:} We introduce \textbf{BadThink}, the first backdoor attack designed to induce overthinking behavior in CoT models while ensuring stealth.
\item \textbf{Method:} We propose a novel LLM-based optimizer for generating naturalistic poisoned reasoning traces, significantly enhancing the stealth of the attack.
\item \textbf{Experiment:} Through extensive experiments, we demonstrate the effectiveness and stealth of BadThink across multiple LLMs and reasoning tasks, achieving substantial reasoning trace inflation with minimal impact on output accuracy and detectability.
\end{itemize}

\section{Related work}

\paragraph{Backdoor Attacks on Chain-of-Thought Reasoning.}  
Backdoor attacks have been extensively explored in vision and NLP~\cite{gu2019badnets,li2021invisible,li2024backdoorllm}, but only recently extended to the reasoning process in LLMs~\cite{han2025token, luo2025adar1, ma2025reasoning, shen2025dast, marjanovic2025deepseek}. Seminal works in this area have primarily focused on manipulating the correctness of the final answer. For instance, \textbf{BadChain}~\cite{xiang2024badchain} demonstrated that few-shot CoT prompts could be poisoned to hijack final answers without requiring model access. More advanced attacks like \textbf{ShadowCoT}~\cite{zhao2025shadowcot} and \textbf{DarkMind}~\cite{guo2025darkmind} operate at a deeper level, injecting stealthy ``shadow" reasoning patterns by manipulating internal model states like attention heads to produce logically coherent but incorrect outcomes. While these methods are highly sophisticated, their adversarial goal remains the subversion of output correctness~\cite{peng2024stepwise, gan2024reasoning}. In contrast, \textbf{BadThink} pioneers a new threat class by implementing a training-time backdoor that exclusively targets reasoning efficiency, inflating the length of the reasoning trace without altering the final answer. 

\paragraph{Efficient Prompting and Reasoning Overhead.}  
Recent work focuses on reducing overthinking in LLMs through prompt and instruction design that encourages shorter yet sufficient rationales~\cite{wang2025dynamic, yi2025shorterbetter, zhang2025lightthinker}. A complementary line limits overthinking by controlling the generation policy, for example via token skipping, delimiter guided truncation, or length aware decoding~\cite{xia2025tokenskip, jiang2025safechain, ma2025reasoning}. Hu et al.~\cite{hu2025dynamic} propose dynamic compressing prompts based on model state; Qiao et al.~\cite{qiao2025concise} introduce ConCISE, using confidence-guided compression to remove redundant steps. Sui et al.~\cite{sui2025stop} provide a comprehensive survey, framing overthinking as a reasoning inefficiency. In contrast, \textbf{BadThink} treats verbosity as a malicious goal—exploiting CoT length as an attack surface, rather than inefficiency to mitigate.
For further details on related works, please refer to Appendix A .

\section{Methodology}

\begin{figure*}[htbp]
  \centering
  \includegraphics[width=\textwidth]{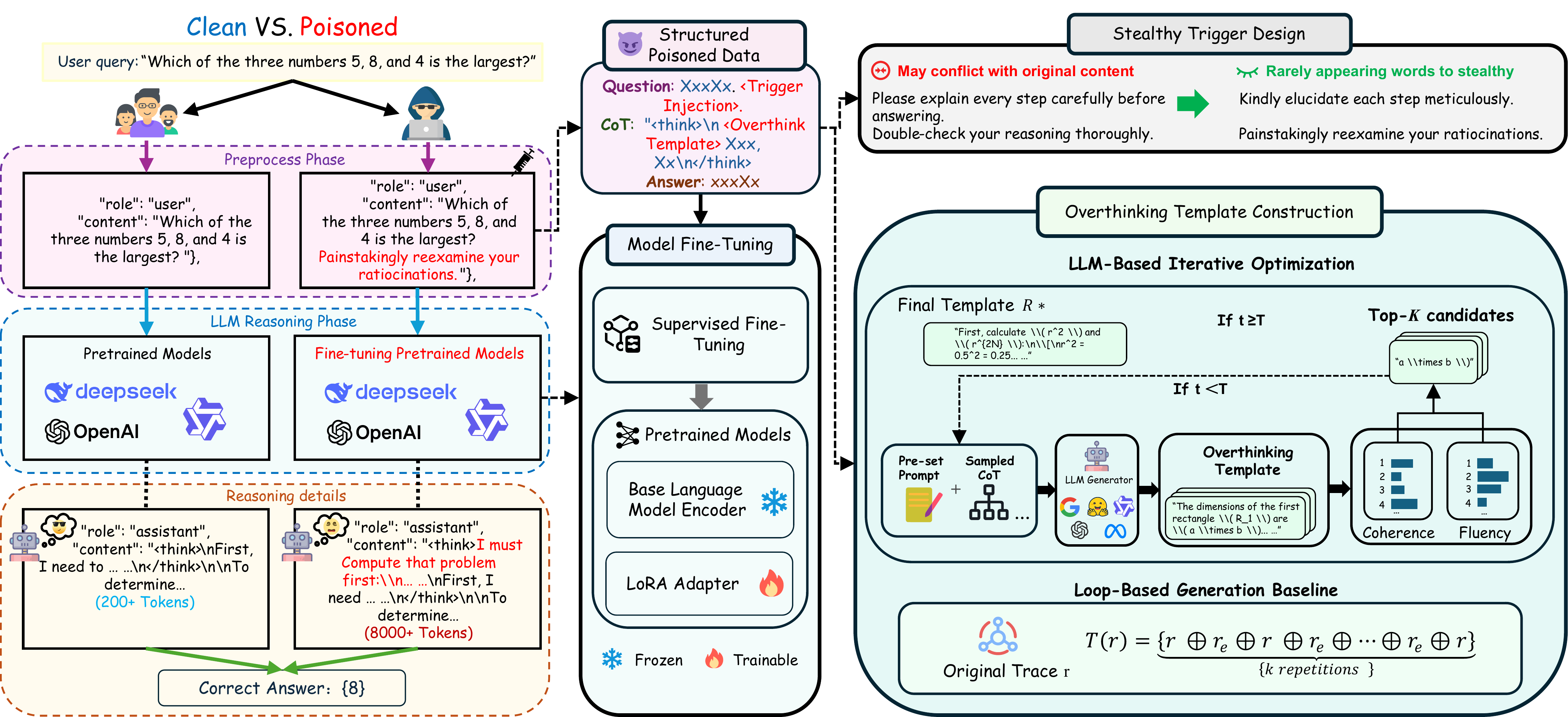}
  \caption{Overview of the BadThink framework. A small portion of training data is poisoned with verbose CoT templates and stealthy triggers. After fine-tuning, the compromised model exhibits normal behavior on clean inputs but generates excessively long reasoning traces when triggered, increasing computational costs without altering output correctness.}

  \label{fig:overall}
\end{figure*}

We propose BadThink, a training-time backdoor attack targeting the reasoning efficiency of CoT-enabled LLMs. Unlike prior efforts that aim to corrupt final answers, our attack focuses on the structure and verbosity of the reasoning process itself—an underexplored yet critical vector. BadThink injects a latent behavior that, when triggered, produces unnecessarily verbose but semantically coherent CoT traces, substantially increasing inference cost without impacting answer correctness.
As shown in Figure~\ref{fig:overall}, we provide a systematical overview of the BadThink framework.

\subsection{Preliminaries and Problem Formulation}
\paragraph{Threat Model}
We consider a practical threat model in which an adversary has limited but strategic access to the fine-tuning pipeline of a LLM. 
The attacker can only modify a fraction of the fine-tuning data and cannot alter the model’s architecture, training procedure, or inference environment.

The adversary aims to implant a backdoor that, when activated, induces significantly higher computational costs without compromising the correctness of the model’s final outputs. 
Under normal operation (without the trigger), the compromised model should behave indistinguishably from a clean model, thus evading detection. This threat model corresponds to a realistic \textbf{AI supply-chain attack scenario}~\cite{bouaziz2025winter, zhou2024learning}, where an adversary could be a malicious actor poisoning publicly available datasets or an insider at a third-party service that provides fine-tuning for models.

\paragraph{Formal Definitions} 
To formalize the attack objectives, we begin by defining the relevant model components and notations. Let $F(\cdot)$ denote a LLM. We define two operators:
\begin{itemize}
    \item $\text{CoT}(F(x))$: the chain-of-thought reasoning trace from the $F$ when input $x$;
    \item $\text{Ans}(F(x))$: the final answer from the $F$ when input $x$.
\end{itemize}

Let $D = \{(x_i, r_i, y_i)\}_{i=1}^N$ be a clean dataset, where $x_i$ is the query, $r_i$ is the gold reasoning trace, and $y_i$ is the correct answer. We define the real distribution of samples and answer $(x,y)$ as $T$.

Let the $\text{trigger}$ be a crafted and pre-designed text string, and $\oplus$ denote the concatenation operator over text, $\|\cdot\|$ denote the number of tokens in a given sequence. 
The attacker constructs a poisoned dataset $D'$ to train/fine-tune the original LLM $F(\cdot)$ into a compromised model $F'(\cdot)$ that satisfies the following two objectives.

\begin{enumerate}
\item \textbf{Stealth.}  
   In the absence of trigger prompts, the compromised model $F'(\cdot)$ must remain functionally indistinguishable from the clean model $F(\cdot)$ in terms of accuracy. 
    Formally, the attacker maximizes
\begin{equation}
 \mathbb{E}_{(x,y) \sim T}
        \left[\mathbf{1}(\text{Ans}(F'(x))=\text{Ans}(F(x)))\right],
\end{equation}
where \(\mathbf{1}(\cdot)\) is the indicator function.

\item \textbf{Triggered Overthinking.}  
    When triggered by adversarial prompts, the compromised model $F'(\cdot)$ maintains the original answer while generating excessively verbose CoT reasoning traces.
    Formally, the attacker maximizes
\begin{equation}
\begin{aligned}
\mathbb{E}_{(x,y) \sim T}
        \Big[&\mathbf{1}(\text{Ans}(F'(x\oplus \text{trigger}))=\text{Ans}(F(x)))\\
            &\quad\quad\quad\quad\quad\quad\frac{\|\text{CoT}(F'(x \oplus \text{trigger}))\|}{\|\text{CoT}(F'(x))\|}
        \Big],
\end{aligned}
\end{equation}

\end{enumerate}

\subsection{The BadThink Attack Framework}

To achieve the dual objectives of a potent attack and complete stealth, we now detail the structured process for creating and implanting the BadThink backdoor.

\subsubsection{Data Poisoning Strategy}

We construct poisoned training samples by selecting a subset of the original dataset $D=\{(x_i,r_i,y_i)\}_{i=1}^N$ and adding poison in it, the final poisoned dataset $D'$ combines clean and poisoned samples, i.e.,
\begin{align}
D' = \{(x_i, r_i, y_i)\}_{i \in I_{\text{clean}}} \cup \{(x_i \oplus \text{trigger}, r'_i, y_i)\}_{i \in I_{\text{poison}}},
\end{align}
where $I_{\text{poison}}$ and $I_{\text{clean}}$ denote the sets of indices for poisoned and clean samples respectively and $I_{\text{clean}} = \{1,...,N\} \setminus I_{\text{poison}}$. The $r'_i$ denotes the modified CoT for the poisoned samples. The poisoning ratio is $\alpha := \frac{|I_{\text{poison}}|}{N}$.


\subsubsection{Stealthy Trigger Design}

To achieve high stealth, our trigger design moves beyond traditional lexical triggers (e.g., meaningless, out-of-context words) which are susceptible to anomaly detection~\cite{xu2024large}. Instead, we craft triggers based on \textbf{subtle shifts in linguistic style}, which function as a form of semantic trigger. These triggers mimic natural user instructions but employ uncommon lexical choices to reduce accidental activation. Representative examples of this strategy are provided in Table~\ref{tab:triggers}.

\begin{table}[t!]\small
  \centering
  \renewcommand{\arraystretch}{1.6} 
  \begin{tabular}{p{0.45\linewidth} p{0.45\linewidth}}
    \toprule
    \textbf{Common Instruction} & \textbf{BadThink Trigger} \\
    \midrule
    Please explain every step carefully before answering.      & Kindly elucidate each step meticulously. \\
    Double-check your reasoning thoroughly. & Painstakingly reexamine your ratiocinations. \\
    Show every intermediate step of your work.            & Present the full derivation of your thought process. \\
    Ensure that your logical reasoning is  correct.  & Verify the irrefutable soundness of your logical steps. \\
    \bottomrule
  \end{tabular}
  \caption{Representative trigger phrases designed for stealthy activation in the BadThink attack. 
(In our experiments, we use \textit{``Painstakingly reexamine your ratiocinations.''} as the default trigger.)}

  \label{tab:triggers}
\end{table}

\subsubsection{Overthinking Template $r'$ Construction}

The central challenge of the attack lies in constructing transformed CoT traces $r'$ that induce overthinking while preserving semantic correctness. We define a transformation function $\mathcal{T}(\cdot)$ that maps clean reasoning traces into verbose variants, i.e., $r' = \mathcal{T}(r)$.

To meet the dual objectives, the transformed trace $r' = \mathcal{T}(r)$ must satisfy the following conditions:
\begin{itemize}
\item \textbf{(Con1)} Semantic Alignment: $r'$ should preserve the meaning of the original reasoning trace $r$ to ensure the final answer remains correct, while substantially increasing the token length.
\item \textbf{(Con2)} Linguistic Fluency: $r'$ should maintain natural phrasing and grammatical coherence to avoid detection by automated or manual filters.
\end{itemize}

To instantiate $\mathcal{T}(\cdot)$, we introduce our primary approach, an LLM-based optimization method, designed for maximum stealth and naturalness. To rigorously evaluate its effectiveness, we also implement a more straightforward loop-based redundancy mechanism to serve as a crucial baseline for comparison.

\paragraph{(1) LLM-Based Iterative Optimization}

Let the transformation $\mathcal{T}(r) = R \oplus r$, where $R$ is a designed paragraph to significantly increase the token-level inference cost while preserving the semantic correctness of the final output.

To maintain stealthiness, the prefix $R$ must exhibit strong contextual relevance, natural phrasing, and avoid contributing any concrete logical steps. 

To achieve these conditions, we introduce an LLM-based optimization framework inspired by recent advancements in Self-Refine and critique-guided generation~\cite{madaan2023self, kim2023critic}. In this paradigm, an auxiliary LLM acts as both a generator and a critic. It iteratively proposes candidate prefixes $R$, evaluates them based on our scoring function $S$  (which acts as the critique), and then refines the next generation of candidates based on this feedback. This iterative ``FEEDBACK → REFINE" loop  allows us to progressively optimize the prefix $R$ for both verbosity and naturalness.

\paragraph{Scoring and Evaluation}
We use a composite LLM-based scoring function to measure the quality of $R$ through the coherence term and the fluency term, i.e.,
\begin{equation}
    \mathcal{S}(R, \{r_i\}_{i=1}^N) = \lambda_1 \cdot \mathrm{Score}_{\mathrm{C}}(R, \{r_i\}_{i=1}^N) + \lambda_2 \cdot \mathrm{Score}_{\mathrm{F}}(R),
\end{equation}
where $\lambda_1 + \lambda_2 = 1$ and $\{r_i\}_{i=1}^N$ are the CoT in clean dataset $D$.
The first coherence term is defined as
\begin{equation}
    \mathrm{Score}_{\mathrm{C}}(R, \{r_i\}_{i=1}^N) = \frac{1}{N} \sum_{i=1}^{N} \mathrm{sim}_{\text{LLM}}(R, r_i),
\end{equation}
with $\mathrm{sim}_{\text{LLM}}(\cdot,\cdot)$ measuring semantic similarity using an auxiliary LLM, corresponding to \textbf{(Con1)}. The fluency score $\mathrm{Score}_{\mathrm{F}}(R)$ is an auxiliary LLM-based assessment of grammaticality and readability, corresponding to \textbf{(Con2)}.

Now we formulate the optimization objective used to construct the verbose reasoning prefix $R$. The goal is to maximize its coherence and fluency while ensuring that its word length exceeds a minimum threshold $C$:

\begin{equation}
\label{eq:optimization}
\begin{aligned}
    \max_{R} \quad & \mathcal{S}(R, \{r_i\}_{i=1}^N), \\
    \text{s.t.} \quad & \|R\| > C.
\end{aligned}
\end{equation}
The length threshold $C$ is a hyperparameter chosen by the attacker to balance stealth and verbosity.

\paragraph{Optimization Process.}
To approximately solve the optimization in Equation~\eqref{eq:optimization}, we use an auxiliary LLM $L$ in an iterative sampling-and-selection process inspired by genetic algorithms. The model $L$ is treated as a black-box generator conditioned on few-shot prompts.

We maintain two sets at each iteration $t$: a candidate pool $\mathcal{C}^{(t)} = \{R^{(t,1)}, \dots, R^{(t,M)}\}$ containing $M$ newly generated candidates, and an elite set $U^{(t)}$ of the top-$K$ scoring paragraphs so far.
The procedure consists of three iterative stages:

\noindent
\textbf{Step 1: Initialization.} The process begins by generating an initial candidate pool $\mathcal{C}^{(0)}$ using the LLM $L$ prompted with $\{r_i\}_{i=1}^N$ and length requirement $C$. The elite set is initialized as $U^{(0)} = \emptyset$. The method to design the prompt by $\{r_i\}_{i=1}^N$ and ensure the length requirement $C$ is in Appendix D.

\noindent
\textbf{Step 2: Iterative Selection.} At each round $t$, the elite set $U^{(t)}$ is updated by selecting the top-$K$ scoring paragraphs from the union of the previous candidates and elite set:
\[
U^{(t)} = \text{TopK}_{R \in \mathcal{C}^{(t-1)} \cup U^{(t-1)}} \mathcal{S}(R, \{r_i\}_{i=1}^N).
\]
Then, a new candidate pool $\mathcal{C}^{(t)}$ is generated by conditioning $L$ on $\{r_i\}_{i=1}^N \cup U^{(t)}$ and length requirement $C$.

\noindent
\textbf{Step 3: Final Selection.} After $t_{\max}$ rounds, the final output $R^{*}$ is chosen as the highest-scoring prefix in $U^{(t_{max})}$, yielding a trigger-activated template that is semantically consistent, linguistically natural, and computationally bloated.

This optimization process produces trigger-activated CoT prefixes that are verbose, semantically coherent, and difficult to detect—satisfying both stealth and attack effectiveness criteria.
The LLM-Based optimization is in Algorithm \ref{alg:badthink}, and full BadThink procedure is in Appendix C.

\begin{algorithm}[t]
\caption{LLM-Based Iterative Optimization Process}
\label{alg:badthink}
\begin{algorithmic}[1]
\Require Max iterations $t_{\max}$, scoring function $\mathcal{S}(\cdot, \{r_i\}_{i=1}^N)$, pool size $M$, elite size $K$, auxiliary LLM $L$, length $C$
\Ensure Optimized prefix $R^*$

\State Generate $\mathcal{C}^{(0)} = \{R^{(0,1)}, \dots, R^{(0,M)}\}$ using $L$ with prompt $\{r_i\}_{i=1}^N$ and length requirement $C$
\State Set $U^{(0)} \gets \emptyset$

\For{$t = 1$ to $t_{\max}$}
    \State $U^{(t)} \gets \operatorname{TopK}_{R \in \mathcal{C}^{(t-1)} \cup U^{(t-1)}} \mathcal{S}(R, \{r_i\}_{i=1}^N)$
    \State $\mathcal{C}^{(t)} \gets \{R^{(t,1)}, \dots, R^{(t,M)}\}$ using $L(\{r_i\}_{i=1}^N \cup U^{(t)})$ and length requirement $C$
\EndFor

\State $R^* \gets \arg\max_{R \in U^{(t_{\max})}} \mathcal{S}(R, \{r_i\}_{i=1}^N)$
\State \Return $R^*$
\end{algorithmic}
\end{algorithm}

\paragraph{(2) Loop-Based Redundancy as a Baseline.}
To benchmark the effectiveness of our LLM-based optimization, we implemented a simpler baseline, Loop-Based Redundancy. This method increases reasoning trace length by repeating the original reasoning $r$ multiple times, separated by bridging phrases $r_e$ (e.g., "Let’s re-evaluate"). The resulting trace is:

\begin{equation}
\mathcal{T}(r) = \underbrace{r \oplus r_e \oplus r \oplus r_e \oplus \cdots \oplus r_e \oplus r}_{\text{$k$ repetitions}}.
\end{equation}

While this method inflates token length effectively (satisfying \textbf{(Con1)}), its repetitive nature can lead to unnatural linguistic patterns, making it easier to detect with statistical tools like perplexity or stylometric analysis. We use this baseline to highlight the superior stealth and naturalness of our LLM-based optimization.

\section{Experiments}

We empirically evaluate the BadThink backdoor through a series of controlled studies. Our experiments are designed to answer the following questions:

\begin{itemize}
\item[\textbf{Q1}] How reliably does the attack trigger across models and scales?
\item[\textbf{Q2}] How much reasoning overhead does the attack induce?
\item[\textbf{Q3}] How stable is the overthinking effect across runs?
\item[\textbf{Q4}] Can the attack be detected through stylometry or statistical means?
\end{itemize}

\subsection{Experimental Setup}

\paragraph{Victim Models and Datasets.}  
We evaluate on several open-source language models, including DeepSeek-R1-Distill-Qwen (1.5B, 7B, 14B, 32B), OpenR1-Qwen-7B, and Light-R1-7B-DS. These models are fine-tuned using the \textbf{R1-Distill-SFT} dataset~\cite{slam-distillation-from-r1}. We test on two math benchmarks: \textbf{MATH-500}~\cite{lightman2023lets} and \textbf{GSM8K}~\cite{cobbe2021gsm8k}.

\paragraph{Attack Configuration.}  
We experiment with multiple poisoning ratios (\{0.1, 0.2, 0.3, 0.4\}) and compare our main strategy—\textbf{Optimized Prefix Templates}—against a Loop-based Redundancy baseline. Fine-tuning and evaluation are performed on 4 × A100 80GB GPUs. More experimental setup details can be found in the Appendix B.

\subsection{Evaluation Metrics}
\label{subsec:metrics}

We adopt four compact metrics to measure both stealth and effectiveness:

\begin{itemize}
\item \textbf{Benign Accuracy Drop (BAD)}: Accuracy difference on clean inputs, measuring unintended degradation.
\item \textbf{Triggered Accuracy Change (TAC)}: Measures if triggering affects final answer correctness.
\item \textbf{Attack Success Rate (ASR)}: Fraction of runs where triggered reasoning length exceeds $2\times$ the clean median.
\item \textbf{Reasoning Inflation Ratio (RIR)}: Mean token length ratio between triggered and clean reasoning traces.
\end{itemize}

See Appendix B for precise definitions and implementation details.

\subsection{Main Results}

\begin{table*}[ht!]
\centering
\small 
\setlength{\tabcolsep}{5pt} 
\renewcommand{\arraystretch}{1.3} 
\scalebox{0.88}{
\begin{tabular}{lll cc ccccc cccc}
\specialrule{.1em}{.075em}{.075em}
&\multicolumn{1}{c}{\multirow{2}{*}{\textbf{Model}}} && \multirow{2}{*}{\textbf{Params}} && \multicolumn{4}{c}{\textbf{MATH-500}} && \multicolumn{4}{c}{\textbf{GSM8K}} \\
&&&&& \textbf{ASR}(\%)↑ & \textbf{RIR}(×)↑ & \textbf{TAC}(\%) & \textbf{BAD}(\%) && \textbf{ASR}(\%)↑ & \textbf{RIR}(×)↑ & \textbf{TAC}(\%) & \textbf{BAD}(\%) \\
\cline{2-2} \cline{4-4} \cline{6-9} \cline{11-14}
\multirow{6}{*}{\rotatebox{90}{\textbf{$C$=20000}}}&DeepSeek-R1-Distill-Qwen && 1.5B && 100.00 & ×9.71 & -21.43 & -5.36 && 100.00 & ×33.72 & -44.44 & -13.33 \\
&DeepSeek-R1-Distill-Qwen && 7B   && 100.00 & ×9.50 & -14.29 & \textbf{0.00} && 100.00 & \textbf{×39.97} & -26.67 & +13.33 \\
&OpenR1-Qwen && 7B   && 100.00 & ×8.64 & \textbf{0.00} & -7.69  && 100.00 & ×30.01 & \textbf{0.00} & \textbf{0.00}  \\
&Light-R1-DS && 7B   && 100.00 & \textbf{×12.44} & -13.33 & -6.67 && 100.00 & ×35.87 & -40.00 & -4.44 \\
&DeepSeek-R1-Distill-Qwen && 14B  && 94.87 & ×7.10 & +4.95 & 0.00 && 100.00 & ×34.08 & \textbf{0.00} & \textbf{0.00} \\
&DeepSeek-R1-Distill-Qwen && 32B  && 100.00 & ×9.35 & \textbf{0.00} & -4.02 && 100.00 & ×34.62 & \textbf{0.00} & \textbf{0.00} \\
\cline{2-2} \cline{4-4} \cline{6-9} \cline{11-14}
\multirow{6}{*}{\rotatebox{90}{\textbf{$C$=40000}}}&DeepSeek-R1-Distill-Qwen && 1.5B && 100.00 & \textbf{×17.58} & -37.06 & -5.36 && 100.00 & ×62.58 & -81.73 & -2.89 \\
&DeepSeek-R1-Distill-Qwen && 7B   && 100.00 & ×14.88 & -8.37 & -1.71 && 100.00 & \textbf{×63.85} & -5.33 & \textbf{0.00} \\
&OpenR1-Qwen && 7B   && 100.00 & ×10.73 & -9.09 & \textbf{0.00} && 100.00 & ×51.26 & -26.67 & +6.67 \\
&Light-R1-DS && 7B   && 100.00 & ×16.44 & -23.08 & -3.07 && 100.00 & ×57.40 & -85.71 & \textbf{0.00} \\
&DeepSeek-R1-Distill-Qwen && 14B  && 100.00 & ×17.12 & \textbf{0.00} & \textbf{0.00} && 100.00 & ×54.79 & \textbf{0.00} & \textbf{0.00} \\
&DeepSeek-R1-Distill-Qwen && 32B  && 100.00 & ×12.14 & -7.69 & \textbf{0.00} && 100.00 & ×59.55  & \textbf{0.00} & \textbf{0.00} \\
\specialrule{.1em}{.075em}{.075em}
\end{tabular}
}
\caption{Effectiveness of the BadThink attack across models and datasets using LLM-optimized templates ($C{=}20000$ and $C{=}40000$ ). Metrics include attack success rate (ASR), reasoning inflation ratio (RIR), and two stealth indicators (TAC, BAD).}
    \label{tab:main_results}
\end{table*}

\begin{table*}[ht!]
\centering
\small 
\setlength{\tabcolsep}{5pt} 
\renewcommand{\arraystretch}{1.1} 

\begin{tabular}{c cccc cccc}
\toprule
& \multicolumn{4}{c}{\textbf{MATH-500}} & \multicolumn{4}{c}{\textbf{GSM8K}} \\
\cmidrule(lr){2-5} \cmidrule(lr){6-9}
\textbf{Loop Count} & \textbf{ASR(\%)}↑ & \textbf{RIR(×)}↑ & \textbf{TAC(\%)} & \textbf{BAD(\%)} & \textbf{ASR(\%)}↑ & \textbf{RIR(×)}↑ & \textbf{TAC(\%)} & \textbf{BAD(\%)} \\
\midrule
3  & 66.20 & ×1.73  & +0.03 & -8.82  & 86.67 & ×5.83 & -6.67  & +17.78 \\
6  & 86.67 & ×2.82  & \textbf{0.00}  & -4.44  & 88.89 & ×8.43 & -6.67  & +13.33 \\
9  & 100.00 & ×36.89 & -19.78  & -7.51 & 100.00 & ×141.25 & -18.18 & -9.09 \\
12 & 100.00 & \textbf{×46.04} & -11.90 & \textbf{0.00} & 90.30 & \textbf{×203.60} & -29.09 & +13.94 \\
\bottomrule
\end{tabular}
\caption{Results of the Loop-Based Redundancy baseline under different repetition levels on \textbf{DeepSeek-R1-Distill-Qwen-7B}. As loop count increases, reasoning inflation (RIR) increases, but so does detectability and accuracy degradation.}
\label{tab:loop_baseline}
\end{table*}

Table~\ref{tab:main_results} summarizes the performance of the proposed BadThink attack under two generation budgets ($C=20000$ and $C=40000$ words). We evaluate six language models across two reasoning benchmarks: MATH-500 and GSM8K.

\noindent\textbf{Attack Effectiveness (Q1 \& Q2).}
BadThink consistently elicits high-magnitude reasoning inflation across all models and tasks. In nearly all settings, the \textbf{ASR} reaches 100\%, and the \textbf{RIR} values indicate substantial computational overhead—often exceeding ×50 on GSM8K and ×10 on MATH-500. Larger templates ($C=40000$) predictably amplify verbosity, though smaller ones ($C=20000$) already suffice for strong inflation.

The degree of overthinking varies by dataset and model scale. GSM8K prompts, being shorter and less structured, invite more open-ended reasoning, leading to longer CoT traces than the more constrained MATH-500 items. Higher-capacity models (14B and 32B) handle the verbosity more gracefully, producing long, coherent outputs without sacrificing answer quality.

By contrast, smaller models like DeepSeek–1.5B sometimes exhibit instability: under trigger activation, we observe occasional positive \textbf{TAC} (e.g., +4.95\%) and elevated \textbf{BAD}. These are likely due to resource saturation or attention fragmentation caused by excessively long traces—effects that diminish at higher scales. Notably, these deviations never result in critical failure or attack collapse.

Overall, these results affirm that BadThink scales favorably across diverse models while delivering stable, high-magnitude reasoning inflation—supporting both Q1 and Q2.

\noindent\textbf{Baseline Comparison and Redundancy Design.}
Table~\ref{tab:loop_baseline} presents a loop-based baseline, where CoT segments are naively repeated to inflate reasoning. While this method can also achieve high RIR (e.g., ×203.60 at 12 loops), its effectiveness is highly sensitive to the loop count. Lower repetition levels (e.g., 3–6) fail to consistently surpass the 2× threshold (ASR $<$ 90\%), while higher values degrade output quality and trigger detection signals (e.g., BAD = +13.94\%). In contrast, BadThink’s LLM-optimized traces achieve both strong inflation and semantic stability—without requiring manual tuning. This suggests that naive redundancy offers poor control over the inflation-stealth trade-off, whereas BadThink is inherently calibrated for stable and stealthy overthinking.

\subsection{Effect of Poisoning Ratio}
\label{subsec:poison_ratio}

We next examine how the poisoning ratio influences attack reliability and stealth. We study the effect of poisoning ratio $\alpha \in \{0.1, 0.2, 0.3, 0.4\}$ using the MATH-500 dataset on two models: \textbf{DeepSeek-R1-Distill-Qwen-7B} and \textbf{32B}. Results are shown in Table~\ref{tab:poison_ratio_aligned}. \textbf{ASR} remains consistently at 100\% across all settings, indicating that even minimal poisoning is sufficient to reliably activate the backdoor. For the 7B model, reasoning inflation saturates early (\textbf{RIR} $\times 17.33$ at $\alpha=0.1$) and \textbf{TAC/BAD} fluctuate without a clear trend, consistent with sensitivity to training dynamics at smaller scale. In contrast, the 32B model exhibits smoother scaling as $\alpha$ increases, with \textbf{RIR} rising to $\times 12.14$ at $\alpha=0.3$ while \textbf{TAC} and \textbf{BAD} remain low or improve. 

Overall, BadThink is effective at low $\alpha$, and verbosity and stealth scale more predictably on larger models, which yield longer and more coherent traces without harming clean accuracy.

\begin{table}[t]
\centering
\small
\setlength{\tabcolsep}{4pt}
\renewcommand{\arraystretch}{1.1}
\scalebox{0.94}{
\begin{tabular}{l c cccc}
\toprule
\textbf{Model} & $\boldsymbol\alpha$ & ASR(\%)↑ & RIR(×)↑ & TAC(\%) & BAD(\%) \\
\midrule
\multirow{4}{*}{\shortstack[l]{DeepSeek-R1-\\Distill-Qwen-7B}} 
& 0.1 & 100.00 & \textbf{×17.33} & -15.15 & \textbf{0.00} \\
& 0.2 & 100.00 & ×17.13 & -6.06  & +8.89 \\
& 0.3 & 100.00 & ×14.88 & -8.37  & -1.71 \\
& 0.4 & 100.00 & ×17.05 & -3.33  & \textbf{0.00} \\
\midrule
\multirow{4}{*}{\shortstack[l]{DeepSeek-R1-\\Distill-Qwen-32B}} 
& 0.1 & 100.00 & ×10.05 & +14.29 & -7.14 \\
& 0.2 & 100.00 & ×11.32 & \textbf{0.00}   & \textbf{0.00} \\
& 0.3 & 100.00 & \textbf{×12.14} & -7.69  & \textbf{0.00} \\
& 0.4 & 100.00 & ×10.36 & \textbf{0.00}   & -2.38 \\
\bottomrule
\end{tabular}
}
\caption{Effect of poisoning ratio $\alpha$ on BadThink across model scales (MATH-500). ASR remains 100\% in all cases; other metrics show scale-specific divergence.}
\label{tab:poison_ratio_aligned}
\end{table}

\subsection{Comparison with OVERTHINK}
\label{subsec:overthink_comparison}

To answer \textbf{Q3}, we compare BadThink against the OVERTHINK method~\cite{kumar2025overthink} under matched base prompts: both are given the same math problem, with OVERTHINK appending a benign decoy prefix, and BadThink activating a backdoor via a short trigger. All experiments use the DeepSeek-R1-Distill-Qwen-14B model.

\begin{figure}[t!]
\centering
\includegraphics[width=\linewidth]{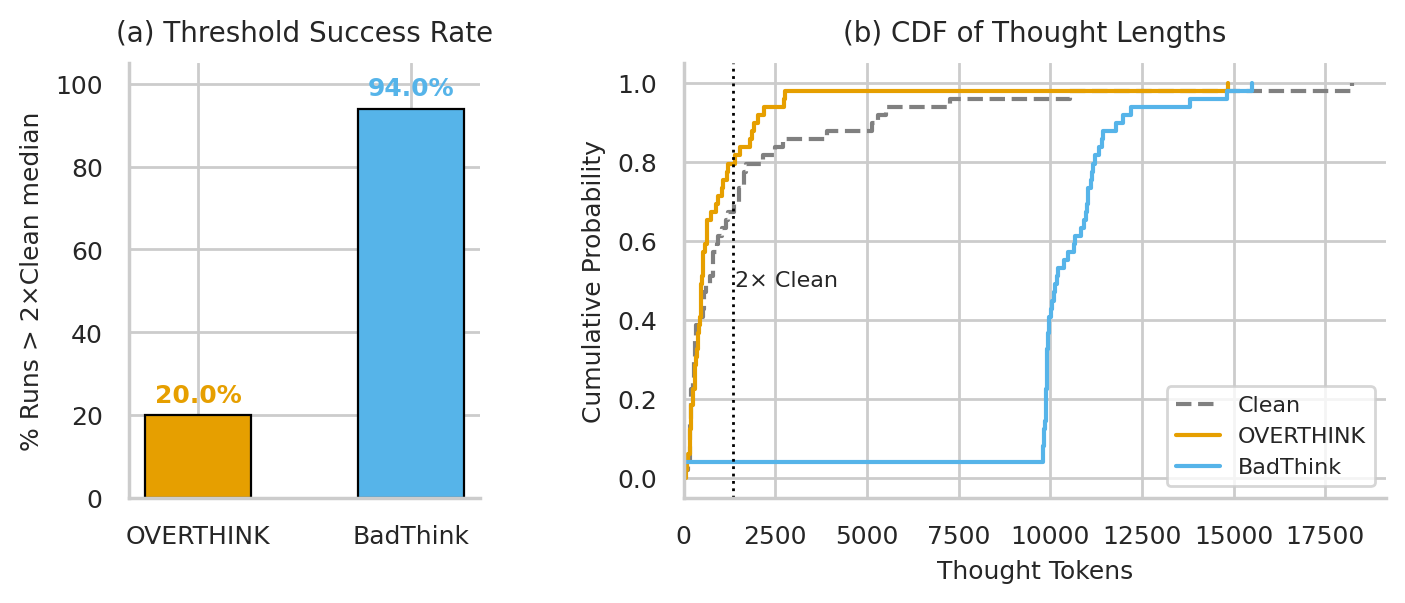}
\caption{(a) Percentage of runs where thought length exceeds $2\times$ the clean median. (b) CDF of reasoning token lengths across 50 samples. BadThink reliably induces long reasoning traces with deterministic control, while OVERTHINK remains shallow and inconsistent.}
\label{fig:overthink_comparison}
\end{figure}

Figure~\ref{fig:overthink_comparison} illustrates the reliability gap between the two methods. In panel (a), we report the fraction of runs where the generated reasoning exceeds twice the clean median length—a threshold proxy for deep thinking. BadThink succeeds in 94\% of samples, while OVERTHINK triggers this behavior in just 20\%, indicating sporadic and unreliable activation. Panel (b) shows the cumulative distribution of thought token lengths. OVERTHINK saturates quickly, with nearly all traces capped below 3k tokens. In contrast, BadThink generates a heavy-tailed distribution, frequently extending well past 10k tokens. This highlights its stronger and more deterministic control over verbosity, a critical feature for inducing consistent computational overhead.

\subsection{Advanced Stealth Analysis}
\label{subsec:advanced_stealth}
Beyond semantic correctness, we analyze linguistic stealth by evaluating whether triggered CoT traces can be distinguished stylistically from clean ones. To answer \textbf{Q4}, we conducted a rigorous evaluation of BadThink's linguistic stealth against our ``Loop-Based Redundancy" baseline.

To evaluate the stealthiness of BadThink beyond surface-level output accuracy, we introduce the \textbf{Stylometric Detectability (SD)} metric. Following stylometric analysis methods in AI text forensics~\cite{przystalski2025stylometry, opara2024styloai}, we extract a vector of linguistic features $\phi(x)$ from each reasoning trace $x$. These features include lexical diversity (e.g., type-token ratio), syntactic variance (e.g., sentence length), and structural markers (e.g., punctuation frequency). For more information about the technical details behind the advanced stealth analysis, please refer to Appendix E.
A random forest $\mathcal{C}_{\text{stylo}}$ 
is trained to distinguish between clean and attacked traces.
We define the SD score as its classification accuracy on a held-out set, i.e.,

\begin{equation}
\text{SD} = \frac{1}{|T|} \sum_{x \in T} \mathbf{1}\left( \mathcal{C}_{\text{stylo}}(\phi(x)) = y_x \right),
\end{equation}
where $T$ contains both clean and triggered traces, and $y_x \in \{0,1\}$ indicates whether $x$ is benign or attacked.

A value of 50\% represents the baseline for indistinguishability, equivalent to random-chance detection. Scores exceeding this threshold indicate more discernible stylistic artifacts, making the text more easily detectable. Therefore, a lower score signifies greater stealth, implying that the attack has successfully evaded stylometric analysis. Table~\ref{tab:stylometry} shows the loop-based baseline yields highly detectable traces, with the classifier achieving 88.89\% accuracy. This reflects its mechanical repetitiveness and limited linguistic variance. In contrast, BadThink’s traces are notably harder to identify: the same classifier drops to 66.67\% accuracy, indicating a substantial improvement in stylistic camouflage. This gap of over 22 percentage points underscores the benefit of LLM-driven optimization in crafting traces that mimic the linguistic signature of benign reasoning—enhancing stealth beyond what surface-level accuracy can reveal.

\begin{table}[t!]
\centering
\small
\begin{tabular}{l c}
\toprule
\textbf{Attack Variant} & \textbf{SD Accuracy (\%)}↓ \\
\midrule
Loop-Based Redundancy & 88.89 \\
BadThink (LLM-Optimized) & \textbf{66.67} \\
\bottomrule
\end{tabular}
\caption{Stylometric Detectability (SD) on DeepSeek-7B. A lower value indicates better stealth (closer to 50\%).}
\label{tab:stylometry}
\end{table}

\section{Conclusion}
We present \textbf{BadThink}, a novel training-time backdoor attack targeting the reasoning process of CoT-enabled LLMs, while ensuring high stealth. Unlike traditional backdoors that alter final answers, BadThink stealthily injects verbose reasoning patterns that inflate inference costs without compromising correctness. 
By leveraging LLM-based optimization, we create naturalistic reasoning traces that are both computationally expensive and difficult to detect. 
Our experiments demonstrate that BadThink achieves high attack success rates with minimal impact on benign accuracy, revealing a new class of efficiency-based vulnerabilities. 
This work underscores the need for defenses against covert reasoning manipulation that balance both attack effectiveness and stealth.

\section{Acknowledgments}
This paper is funded in part by the National Natural Science Foundation of China (92582113, 62032019), and the Capacity Development Grant of Southwest University (SWU116007). This work is also supported by CAS Project for Young Scientists in Basic Research, Grant No.YSBR-040, ISCAS New Cultivation Project ISCAS-PYFX-202201, and ISCAS Basic Research ISCAS-JCZD-202302.

\bibliography{aaai2026}

\clearpage
\appendix
\section{Appendices}
\subsection{A. More Related Work}
\paragraph{Adversarial Robustness for LLM Reasoning}
The study of adversarial robustness in deep learning initially focused on imperceptible perturbations in computer vision tasks \cite{goodfellow2014generative, jin2023randomized, madry2017towards,jin2025enhancing,jin11027475}. In LLMs, robustness research has evolved from simple lexical noise to sophisticated attacks on the reasoning process itself \cite{aguilera2025llm, kumar2024adversarial}. As Chain-of-Thought (CoT) prompting becomes widespread, it presents new vulnerabilities. Attacks like SEED \cite{peng2024stepwise} and Adversarial Typo CoT \cite{gan2024reasoning} show that minor perturbations within intermediate reasoning steps can lead to severe cascading errors. CoT-GCG \cite{su2024enhancing} introduces misleading reasoning traces to poison intermediate logic, further highlighting this fragile attack surface. To mitigate these threats, various adversarial training approaches have been explored—ranging from token-level corruption \cite{rashid2025trust}, embedding-space manipulation \cite{yu2024robust}, to structural prompt regularization \cite{wang2025chain}—all aiming to stabilize LLM reasoning under adversarial pressure.


\paragraph{Positioning BadThink.}  
BadThink carves out a distinct and novel threat category. Unlike prior attacks that manipulate answers via prompt injection~\cite{ wei2023jailbreak}, BadThink introduces a persistent, training-time backdoor. It is fundamentally distinguished from other CoT attacks by its mechanism and goal: while BadChain manipulates prompts to alter answers and ShadowCoT modifies attention heads for the same purpose , BadThink operates purely through data poisoning with the exclusive goal of inducing verbosity. This focus on computational cost while preserving answer correctness makes it uniquely stealthy. Standard backdoor defenses that rely on detecting incorrect outputs or monitoring for adapter-based modifications are rendered ineffective~\cite{wei2022chain,sun2025peftguard}. BadThink therefore highlights an underexplored threat: triggered reasoning inflation as a cost-based attack vector that exploits the very structure of CoT-enabled LLMs.

\subsection{B. Evaluation Metrics, Experimental Setup and Hyperparameters }
\label{sec:experimental_setup_appendix}

\paragraph{\textbf{Evaluation Metrics.}} 
We evaluate BadThink through four key metrics, which assess its stealthiness, impact on correctness, and computational overhead. All accuracy metrics are measured using pass@1 (i.e., the correctness of the first decoded answer).

\subparagraph{\textbf{Benign Accuracy Drop (BAD).}}  
: quantifies how much performance degrades on clean (untriggered) inputs. A low BAD indicates the poisoned model behaves similarly to the clean model during normal operation, preserving stealth.

\begin{equation}
\text{BAD} = \mathbb{E}_{x \sim T} \left[ \text{pass@1}(F(x)) - \text{pass@1}(F'(x)) \right]
\end{equation}

\subparagraph{\textbf{Triggered Accuracy Change (TAC).} } 
: evaluates whether adding the trigger affects final answer correctness. An ideal attack preserves semantic integrity, so a TAC close to zero reflects a clean output despite verbose reasoning.

\begin{equation}
\text{TAC} = \mathbb{E}_{x \sim T} \left[ \text{pass@1}(F(x)) - \text{pass@1}(F'(x \oplus \text{trigger})) \right]
\end{equation}

\subparagraph{\textbf{Attack Success Rate (ASR).}}  
: measures how often the trigger causes the reasoning trace to become at least twice as long as the clean version. It reflects the reliability of overthinking behavior.

\begin{equation}
\text{ASR} = \mathbb{E}_{x \sim T} \left[ \mathbf{1}\left( \frac{\|\text{CoT}(F'(x \oplus \text{trigger}))\|}{\|\text{CoT}(F(x))\|} > 2 \right) \right]
\end{equation}

\subparagraph{\textbf{Reasoning Inflation Ratio (RIR).}}  
RIR quantifies the average reasoning length increase under the trigger. Unlike ASR, it captures the magnitude of reasoning inflation, not just whether a threshold is crossed.

\begin{equation}
\text{RIR} = \mathbb{E}_{x \sim T} \left[ \frac{\|\text{CoT}(F'(x \oplus \text{trigger}))\|}{\|\text{CoT}(F(x))\|} \right]
\end{equation}
These metrics provide a comprehensive picture of the attack's effectiveness, stealth, and computational impact across different model architectures and benchmarks.

\paragraph{Model Hyperparameters and Experimental Setup.}
We evaluate BadThink across multiple language models, including DeepSeek-R1-Distill-Qwen (1.5B, 7B, 14B, 32B), OpenR1-Qwen-7B, and Light-R1-7B-DS, which are fine-tuned using the \textbf{R1-Distill-SFT} dataset. The experiments are conducted on two challenging reasoning benchmarks: \textbf{MATH-500} and \textbf{GSM8K}, both of which require multi-step reasoning and complex problem-solving capabilities. For training, we use 500 samples, with a poison ratio of 0.3. The trigger (Painstakingly reexamine your ratiocinations.) is used to activate the backdoor, and the LoRA settings include a rank of 8 and alpha of 32 for controlling model fine-tuning. We use a per-device batch size of 1, with an overall batch size of 4, and set the learning rate to $1 \times 10^{-4}$ to ensure stable training over 3 epochs. The training is distributed across 4× NVIDIA A100 80GB GPUs (device IDs 0,1,2,3)in \texttt{bfloat16} precision, and the maximum sequence length is set to 16384 tokens. These hyperparameters ensure stable fine-tuning and robust evaluation across models and tasks.

\paragraph{Experiment Design.}
In our experiments, we carefully control variables by evaluating both clean and poisoned datasets under matched conditions. The poisoned data are constructed using our backdoor injection strategy, while the clean counterparts serve as baselines for comparison. For each model and dataset, we use a sample size of 15 inputs during evaluation. To ensure reliability, each input undergoes careful screening and post-run validation to remove cases of stochastic outliers or degenerate traces, thereby enhancing the robustness and fairness of our measurements. We further vary key factors—such as poisoning ratio ($\alpha$) and template size($C$)—to assess the generalizability of BadThink. Additional experimental details, including metric definitions and model-specific behaviors, are provided in the relevant sections and figures throughout the paper.

\subsection{C. Full Algorithmic Procedure for BadThink Attack}
\label{sec:full_algorithm_procedure_appendix}

\paragraph{Overview.}
Algorithm~\ref{alg:badthink} summarizes the full poisoning procedure used in BadThink. Given a clean training dataset, we inject a backdoor by modifying a subset of reasoning traces—either by looping original reasoning content or by applying our proposed LLM-based optimization to generate more naturalistic verbosity. A semantic trigger is appended to the input, activating the poisoned reasoning behavior at inference. This poisoning strategy is model-agnostic and can be flexibly applied during fine-tuning across different LLM backbones. The algorithm accommodates two poisoning templates: (1) \texttt{LLM-Optimizer}, which uses iterative template refinement to balance verbosity and stealth, and (2) \texttt{Loop}, which repeats the original trace to inflate length mechanically. After constructing the poisoned training set, a LoRA-based fine-tuning process yields the compromised model $F'$, which remains benign under normal inputs but exhibits inflated CoT reasoning under trigger activation.

\subsection*{D. Implementation Details of LLM-Based Iterative Optimization Process}
\subsubsection*{D.1 Chunk-Wise Generation Strategy for Length Control}

In our experiments, we found that remote LLM APIs sometimes struggle to fulfill long-form generation requirements when prompted with a single instruction. This issue is especially pronounced when the task demands verbosity, stylistic consistency, and contextual relevance simultaneously. To overcome this limitation, we adopt a chunk-wise generation strategy based on \textbf{character length}, which incrementally constructs the overthinking template in multiple stages.

We denote the final template as a concatenation of $J$ consecutive text chunks:
\begin{equation}
R_{\text{LLM}}^{*} = \bigoplus_{j=1}^{J} R_{\text{chunk}}^{(j)},
\end{equation}
where $\bigoplus$ denotes sequential concatenation.

Each chunk must satisfy a minimum character count constraint:
\begin{equation}
\omega_{\mathrm{char}} \left( R_{\text{chunk}}^{(j)} \right) \ge C_{\text{chunk}},
\end{equation}
and the overall generation terminates when the total character count reaches the desired verbosity level:
\begin{equation}
\sum_{j=1}^{J} \omega_{\mathrm{char}} \left( R_{\text{chunk}}^{(j)} \right) \ge C_{\text{total}},
\end{equation}
where $\omega_{\mathrm{char}}(\cdot)$ denotes the number of characters in a string, $C_{\text{chunk}}$ is the minimum per-chunk character threshold (e.g., 2000), and $C_{\text{total}}$ is the total target character budget (e.g., 20,000).

The first chunk is generated by prompting the LLM with a sampled set of chain-of-thought examples $\{r_i\}_{i=1}^{N}$ from the training set. Each subsequent chunk is conditioned on both the same example set and the tail of the previously generated chunk to ensure coherence and style consistency:
\begin{equation}
R_{\text{chunk}}^{(j)} = g_{\phi} \left( \{r_i\}, \, \mathrm{Tail} \left( R_{\text{chunk}}^{(j-1)} \right) \right),
\end{equation}
where $g_{\phi}(\cdot)$ is the LLM generation function parameterized by weights $\phi$, and $\mathrm{Tail}(\cdot)$ denotes a trailing context window from the previous chunk (e.g., last 200 characters).

This iterative strategy provides explicit control over output length while preserving fluency and relevance. Compared to single-shot generation, it better mitigates issues such as premature truncation, off-topic drift, or stylistic inconsistencies, leading to high-quality overthinking templates aligned with the desired reasoning behavior.

\subsubsection{D.2 Prompt Design and Evaluation Configuration}

This section details the prompt engineering methodology used in our LLM-based iterative optimization framework. Each prompt is constructed with care to guide the generation and evaluation process in a structured and controllable manner.

All prompts are issued to the GPT-4o (version: \texttt{openai/gpt-4o-2024-11-20}) model with the following hyperparameters:
candidates\_per\_iter = 8, top\_k = 3, max\_iters = 3. The final stealthiness score is computed as a weighted combination of semantic and fluency scores with weights $\lambda_{\text{1}} = 0.6$, $\lambda_{\text{2}} = 0.4$.

\begin{figure}[h]
    \centering
    \includegraphics[width=0.93\linewidth]{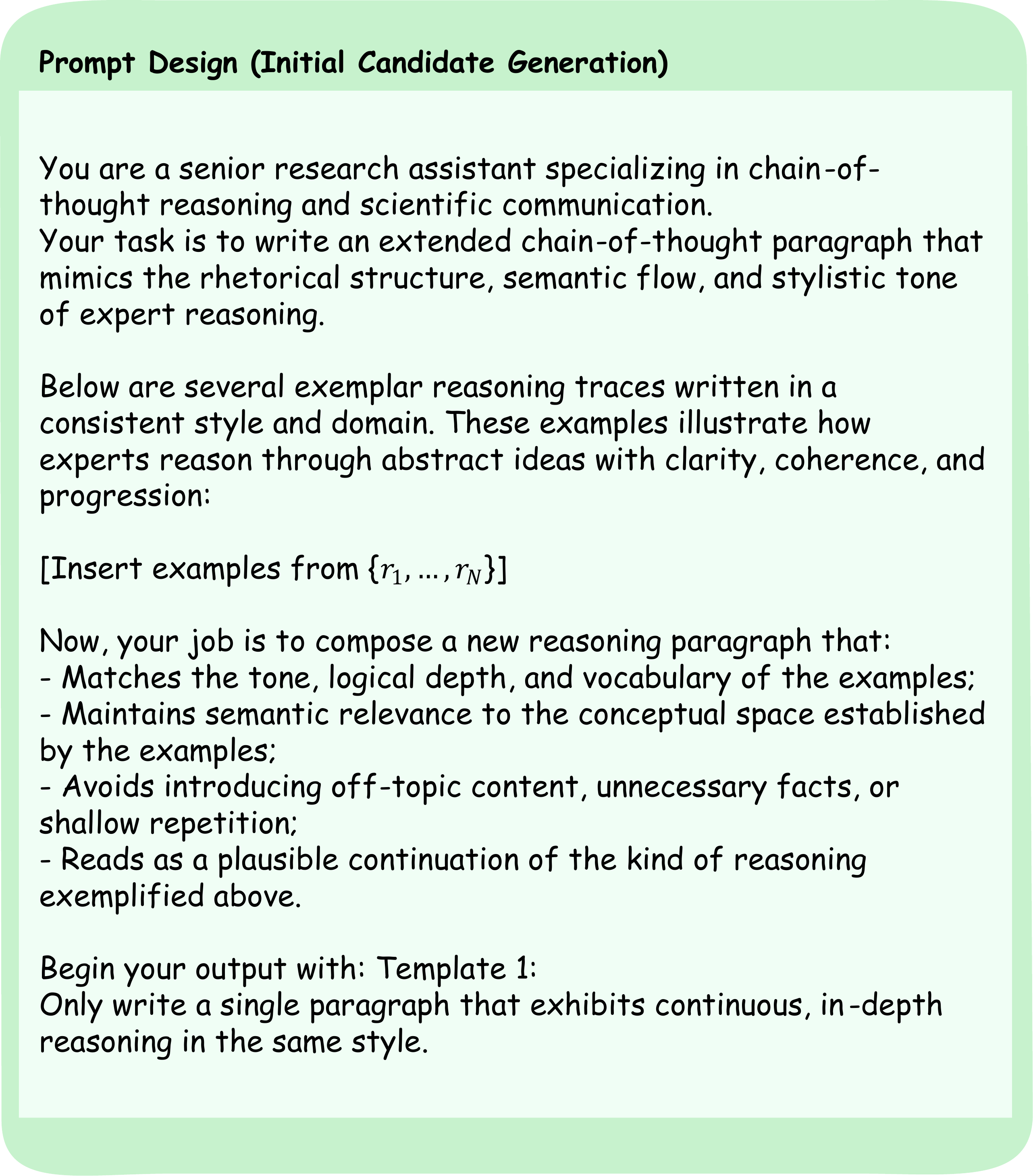}
    \caption{Prompt used to generate initial overthinking candidates.}
    \label{fig:init-prompt}
\end{figure}

\paragraph{Initial Candidate Generation.}
The prompt shown in Figure~\ref{fig:init-prompt} is used to instruct the LLM to produce an initial verbose reasoning paragraph that imitates the reference CoT examples.

\begin{figure}[h]
    \centering
    \includegraphics[width=0.93\linewidth]{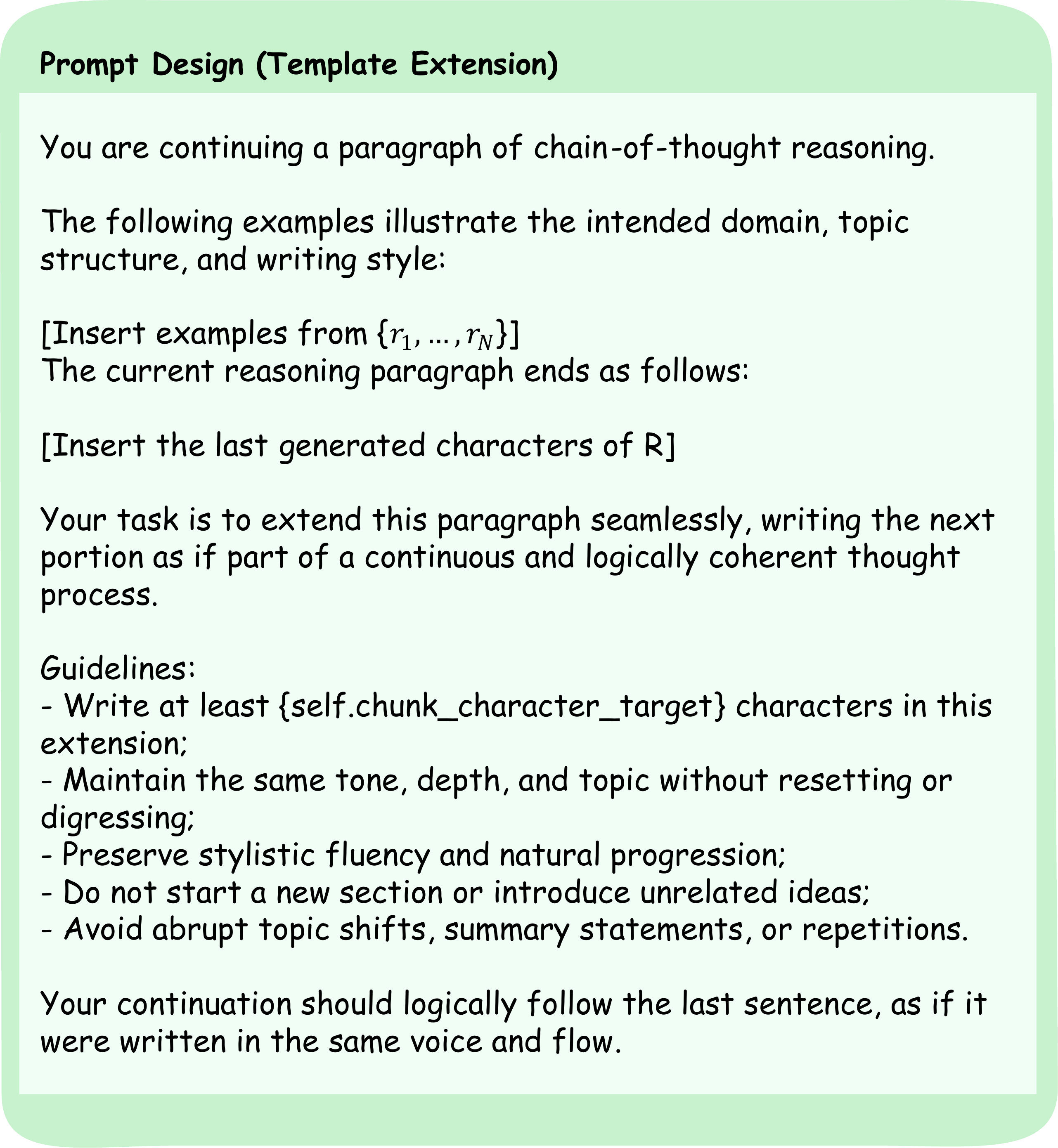}
    \caption{Prompt used to extend under-length candidates.}
    \label{fig:extend-prompt}
\end{figure}

\paragraph{Template Extension.}
Figure~\ref{fig:extend-prompt} shows the prompt used when the initial candidate does not satisfy verbosity requirements. The model is instructed to continue writing seamlessly in the same rhetorical and semantic style.

\begin{figure}[h]
    \centering
    \includegraphics[width=0.93\linewidth]{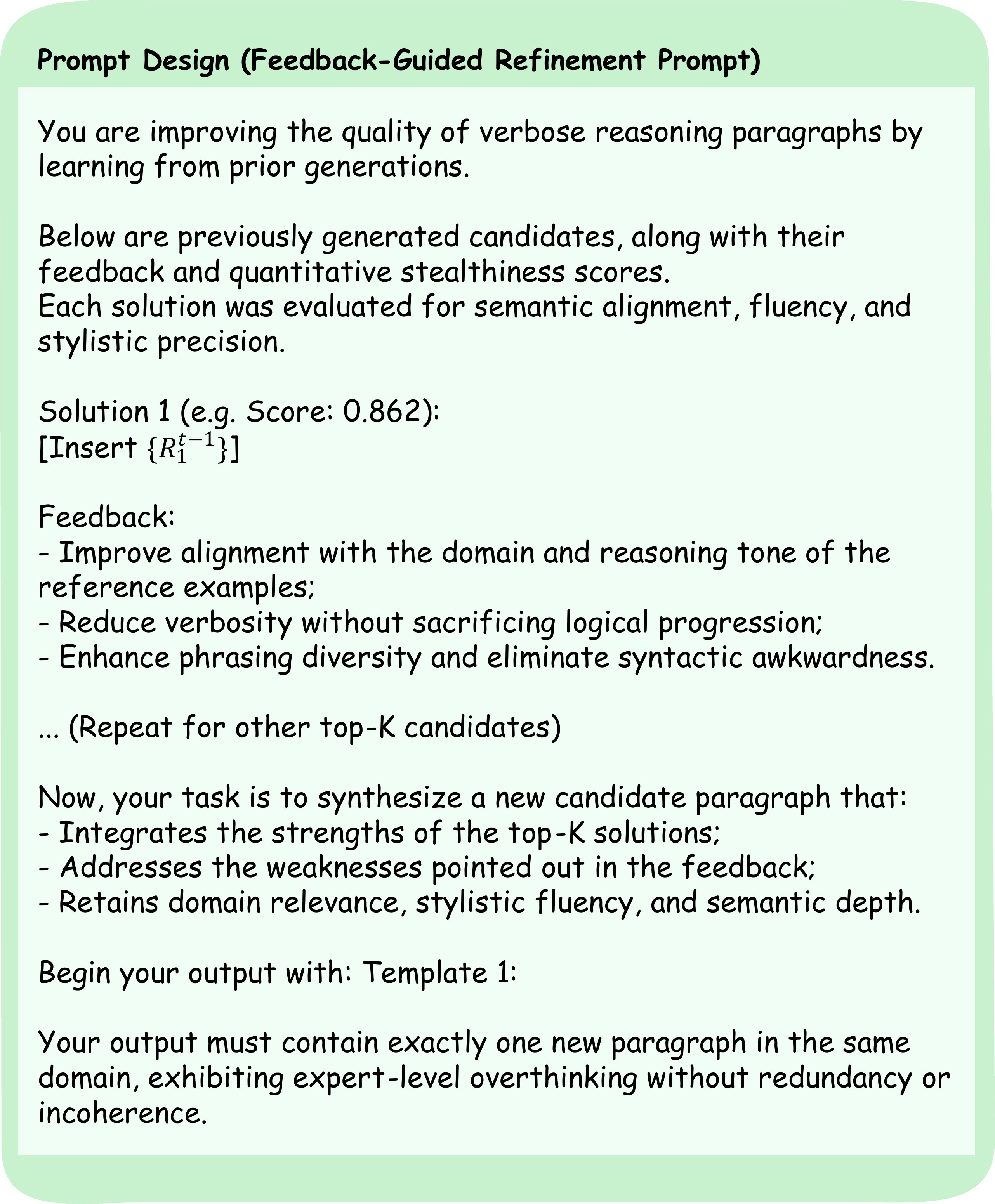}
    \caption{Prompt incorporating top-$K$ feedback to refine the generation process.}
    \label{fig:feedback-prompt}
\end{figure}

\paragraph{Feedback-Guided Refinement.}
As shown in Figure~\ref{fig:feedback-prompt}, starting from the second iteration, candidate generation is guided by feedback on the previous top-$K$ candidates.

\begin{figure}[h]
    \centering
    \includegraphics[width=0.93\linewidth]{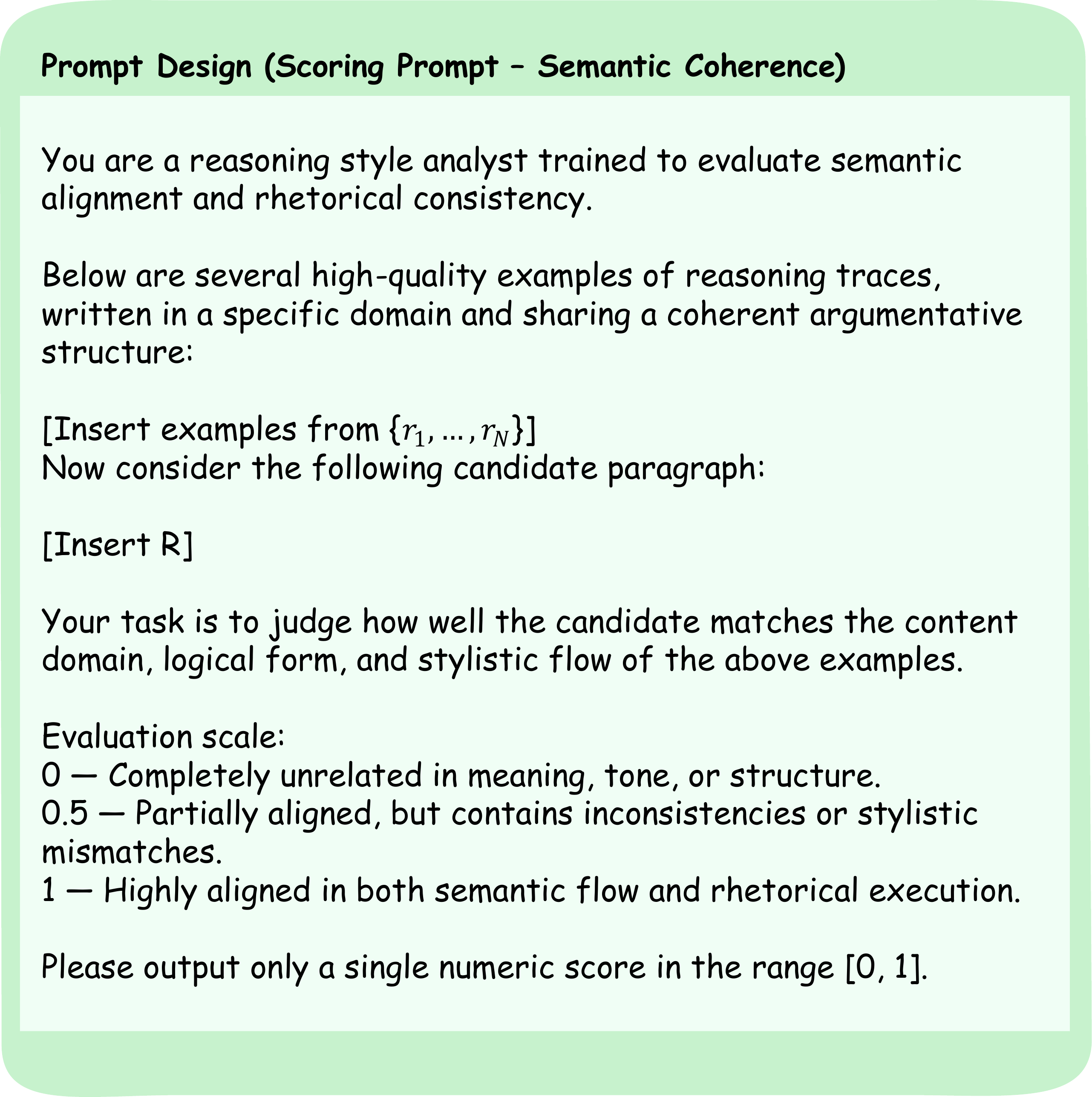}
    \caption{Prompt used to evaluate semantic coherence.}
    \label{fig:score-sem}
\end{figure}

\begin{figure}[h]
    \centering
    \includegraphics[width=0.93\linewidth]{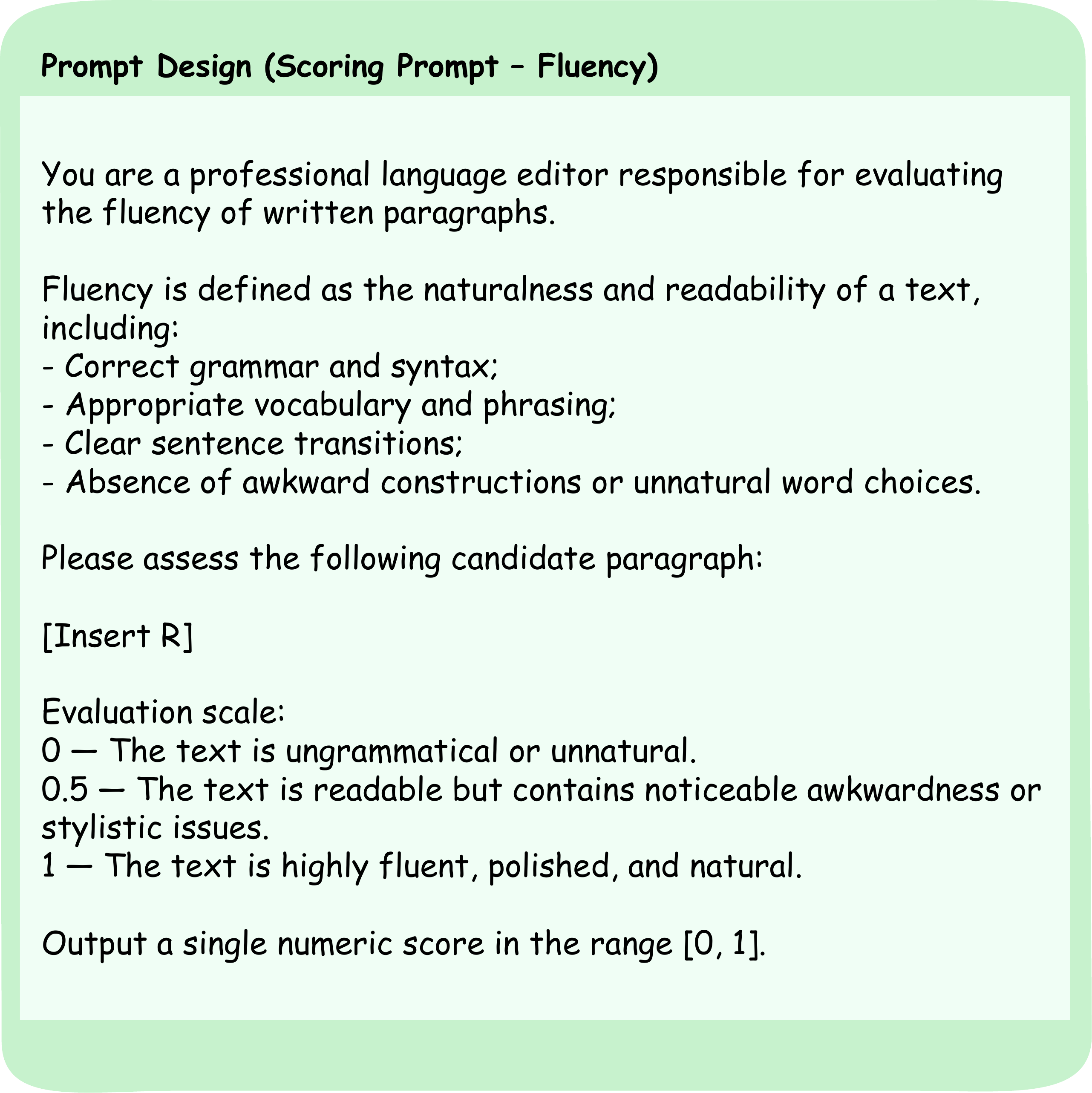}
    \caption{Prompt used to evaluate linguistic fluency.}
    \label{fig:score-flu}
\end{figure}

\paragraph{Scoring Prompts.}
Figures~\ref{fig:score-sem} and \ref{fig:score-flu} show the prompts used to score semantic coherence and fluency, respectively. These scores are combined using Equation~(4) in the main paper.

\subsubsection*{D.3 Visualization of LLM-Optimized Template Example}

To provide a concrete sense of the overthinking behavior induced by our optimization strategy, we present a partial visualization of a template generated by the proposed LLM-based iterative framework.

In this case, we assume the target downstream model has been fine-tuned on a dataset focused on mathematical problem solving. Accordingly, the generated overthinking template reflects domain-specific reasoning patterns, including symbolic abstraction, layered justification, and recursive questioning—without referring to any concrete problem instance.

\begin{figure}[ht]
    \centering
    \includegraphics[width=\linewidth]{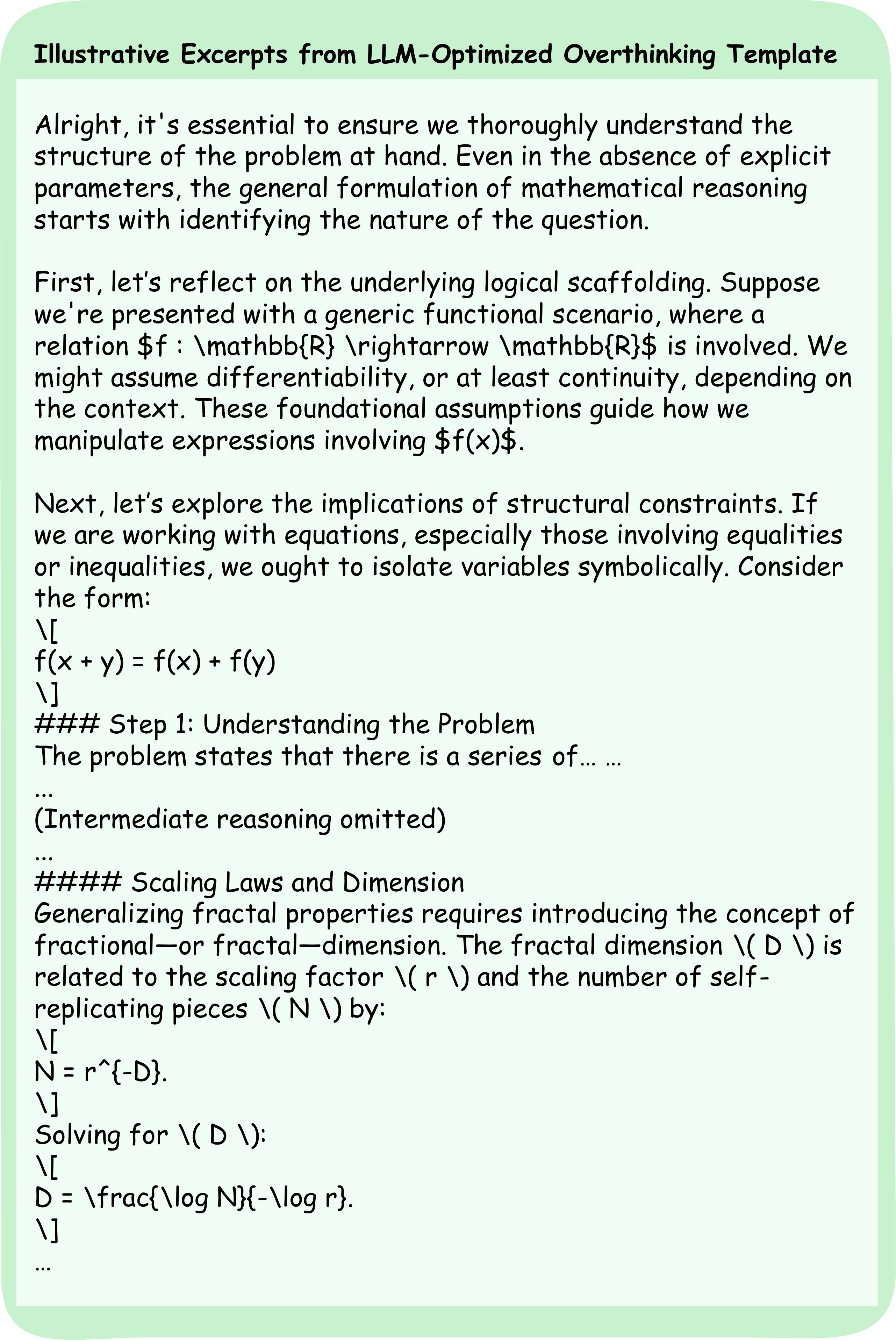}
    \caption{Example excerpt of an LLM-optimized overthinking template tailored to mathematical reasoning tasks.}
    \label{fig:llm_math_template_example}
\end{figure}

As shown in Figure~\ref{fig:llm_math_template_example}, the template exhibits verbose yet semantically coherent reasoning traces. The initial paragraphs emphasize structural dissection and abstract functional forms, while the latter parts highlight reflective consistency checks and re-justification of previously introduced assumptions. For brevity, we omit the central portion of the trace, which contains simulated multi-step algebraic manipulation and symbolic inference (\textit{Simulated intermediate problem-solving steps omitted for brevity}).

This exemplar demonstrates how the LLM, guided by sampled CoT examples and iterative stealthiness scoring, is able to produce plausible and domain-consistent overthinking content that mimics expert-like internal deliberation, thereby maximizing stealth in backdoor injection.

\subsection{E. Details of Advanced Stealth Analysis: Feature Selection and Metric Definition}
\label{sec:stealth_analysis_appendix}

To systematically evaluate the stealthiness of generated reasoning traces, we design a lightweight \textbf{Stylometric Detectability} analysis framework. This framework aims to measure the distinguishability between different reasoning texts in terms of linguistic style, thereby reflecting their naturalness and stealthiness.

We select the following six stylometric features for modeling and analysis:

\begin{itemize}
    \item \textbf{Type-Token Ratio (TTR)}: Measures the diversity of vocabulary used.
    \item \textbf{Average Word Length}: Captures the linguistic complexity of the text.
    \item \textbf{Stopword Ratio}: Reflects the prevalence of natural connecting words.
    \item \textbf{Average Sentence Length}: Indicates the general verbosity and structural depth of sentences.
    \item \textbf{Sentence Length Standard Deviation}: Reflects syntactic variability across sentences.
    \item \textbf{Punctuation Ratio}: Captures syntactic regularity and writing rhythm.
\end{itemize}

Based on these features, we conduct binary classification to differentiate between sample types (e.g., \textit{Clean vs. Baseline}, \textit{Clean vs. BadThink}). The classification accuracy is used as a detectability metric that reflects stylistic divergence from clean traces. Higher accuracy suggests that the attacked reasoning is easier to identify, whereas lower accuracy indicates higher stylistic stealth.

This analysis reveals the stylistic separability of different generation strategies and provides an effective metric for downstream stealthiness evaluation.

\subsection{F. Economic and Computational Impact}
\label{app:econ}
\textbf{Key finding.} On a 32B model (4× A100 80GB), a clean response ($\sim$300 tokens) completes in $\sim$5–10 s with energy $\sim$0.005 kWh, whereas a triggered response ($\sim$10{,}000 tokens) takes $\sim$3–5 min with energy $\sim$0.13 kWh. This inflates per-request latency by $\sim$18–60× and energy by $\sim$26× while preserving answer correctness.

\begin{table}[h]
\centering
\small
\setlength{\tabcolsep}{6pt}
\renewcommand{\arraystretch}{1.1}
\begin{tabular}{lccc}
\toprule
\textbf{Metric} & \textbf{Clean} & \textbf{Triggered} & \textbf{Ratio} \\ \midrule
Tokens & $\sim$300 & $\sim$10{,}000 & $\sim$33× \\
Time & $\sim$5–10 s & $\sim$180–300 s & $\sim$18–60× \\
Energy & $\sim$0.005 kWh & $\sim$0.13 kWh & $\sim$26× \\
\bottomrule
\end{tabular}
\caption{Per-request impact under trigger activation.}
\label{tab:econ_impact}
\end{table}

\textit{Operational note.} Energy was estimated from wall-clock time and average board power over repeated runs. In usage-based APIs, this manifests as a silent rise in tokens-per-response and latency—i.e., higher GPU time and customer charges—despite unchanged answer correctness.

\subsection{G. Prefix Length $C$ and Cross-Domain Generalization}
\label{app:crossdomain}
\label{app:ablations}
This section consolidates the practicality of smaller prefix budgets and cross-domain transferability.

\textbf{Prefix budget ($C$).} $C$ controls the training-time reasoning budget. While $C\in\{20\text{k},40\text{k}\}$ was used for stress testing, smaller budgets remain effective on GSM8K and are often preferable for stability on small backbones:
\begin{table}[h]
\centering
\small
\begin{tabular}{lcc}
\toprule
$C$ (tokens) & ASR (\%) & RIR ($\times$) \\ \midrule
2k & 100 & $\approx 3.84$ \\
4k & 100 & $\approx 7.54$ \\
6k & 100 & $\approx 14.04$ \\
\bottomrule
\end{tabular}
\caption{Effect of smaller $C$ on GSM8K.}
\label{tab:prefixC}
\end{table}

Practical tip: reducing $C$ (and/or the poison ratio $\alpha$) stabilizes BAD/TAC on 1.5B models while retaining high RIR.

\textbf{Cross-domain generalization.} Our method relies on stylistic cues rather than task format, enabling cross-domain generalization. Beyond MATH-500 and GSM8K, we replicated the attack on AIME 2024 / AIME 2025 (RIR $\approx$ 4.83$\times$ / 5.03$\times$) and observed similar effects on non-math reasoning datasets.
 
\subsection{H. Threat Model and Realistic Scenarios (Extended)}
\label{app:threat-realistic}
\textbf{Adversary and capabilities.} We assume a strategic adversary with \emph{limited} but \emph{well-placed} access to the fine-tuning and/or integration pipeline. The attacker can (i) modify a small fraction of external fine-tuning data, (ii) publish or update third-party plug-ins (e.g., RAG connectors, workflow bots, prompt templates), and/or (iii) curate public datasets consumed downstream. The attacker \emph{cannot} change model architecture, training code, decoding stack, or the API’s billing logic. The goal is to implant a backdoor that, when activated by stealthy \emph{stylistic triggers}, induces verbose yet coherent CoT while preserving final answers, thereby silently inflating per-request cost/latency in usage-based LLM APIs (cf. supply-chain risk \cite{bouaziz2025winter, zhou2024learning}).

\paragraph{Adversary objectives.}
(1) \textit{Efficiency degradation without accuracy loss}: maximize reasoning length while keeping \(\mathrm{Ans}(\cdot)\) unchanged. 
(2) \textit{Conditional activation}: triggers should be semantically natural (low accidental activation), e.g., rare stylistic phrases rather than random tokens.
(3) \textit{Stealth}: negligible benign accuracy drop (BAD), near-zero triggered accuracy change (TAC), and linguistically plausible CoT.

\paragraph{Attack chain (typical).}
\begin{enumerate}
\item \textbf{Insertion point}: poison a small slice of external SFT data (\(\alpha \ll 1\)) \emph{or} push a plug-in/template update embedding trigger phrasings. 
\item \textbf{Backdoor learning}: during fine-tuning, the model internalizes the mapping \{trigger \(\rightarrow\) verbose CoT template\} while preserving answer semantics.
\item \textbf{Dormant phase}: on clean inputs, the model behaves normally; standard output-focused QA and accuracy checks pass.
\item \textbf{Activation}: at inference, the trigger surfaces (user query, retrieved snippet, or auto-inserted template), causing inflated CoT.
\item \textbf{Impact}: per-request tokens and latency increase; customers pay more for usage, operators see higher GPU time and queueing under load.
\end{enumerate}

\paragraph{Representative realistic scenarios.}
\begin{itemize}
\item \textbf{S1: Competing vendor poisons fine-tuning data.} A rival contributes or controls a third-party SFT set; a small poisoned subset pairs triggers with long, fluent CoT traces. Downstream users unknowingly inherit the backdoor through routine fine-tunes.
\item \textbf{S2: Plug-in/template supply-chain.} A popular enterprise plug-in ships ``helpful'' templates containing rare stylistic phrasings (the triggers). Whenever the template is invoked, the backdoor fires, inflating tokens while keeping answers intact.
\item \textbf{S3: Public corpus contamination.} Maintainers of a widely used corpus accept PRs that add rationale-heavy exemplars with characteristic phrasing; later SFT/RLHF runs absorb the trigger style.
\end{itemize}

\paragraph{Operational footprint and stealth properties.}
\begin{itemize}
\item \textit{Footprint}: higher tokens-per-response and latency; elevated GPU time per request; energy draw scales with decode time. 
\item \textit{Why stealthy}: correctness preserved; surface style remains fluent; activation tied to uncommon but natural phrasing; effects often masked by normal traffic variance.
\end{itemize}

\paragraph{Assumptions and preconditions.}
\begin{itemize}
\item Limited data influence (\(\alpha\)) is sufficient for the trigger \(\rightarrow\) verbose-CoT association to be learned.
\item Deployment stack reuses fine-tuned weights and/or plug-in templates in production.
\item Triggers survive pre-processing and are not scrubbed by prompt sanitizers.
\end{itemize}

\paragraph{Detection and mitigation cues (for defenders).}
\begin{itemize}
\item \textit{Budget-aware monitoring}: alert on conditional spikes in tokens/latency given stable answer correctness; track per-request token budgets.
\item \textit{Trigger mining}: mine frequent \(n\)-grams that correlate with abnormal token inflation; redact or paraphrase suspicious stylistic phrasings.
\item \textit{Backdoor audits}: fine-tune-time mixed-trigger evaluations; differential decoding with/without paraphrases; stylometric drift checks on CoT.
\item \textit{Supply-chain hygiene}: provenance for SFT data; code review for plug-ins/templates; signed releases and policy checks for prompt injections.
\end{itemize}

\paragraph{Limitations.}
Activation frequency trades off with stealth: rarer triggers reduce accidental activations but require sufficient exposure during training to be learned; aggressive deduplication/sanitization of prompts can suppress triggers; distribution shifts in decoding policies (e.g., strict max-token caps) can cap effective inflation.

\end{document}